\documentclass[journal,11pt,twocolumn]{IEEEtran}
\usepackage{fixltx2e}
\usepackage{cite}
\usepackage{mathrsfs}
\usepackage{color}
\usepackage{float}
\usepackage[caption=false]{subfig}
\usepackage[table, dvipsnames]{xcolor}
\usepackage{array,booktabs}
\usepackage{balance}
\ifCLASSINFOpdf
   \usepackage[pdftex]{graphicx}
   \graphicspath{{Figs/}}
   \DeclareGraphicsExtensions{.pdf,.jpeg,.png}
\else
\fi
\usepackage[cmex10]{amsmath}
\usepackage{amsmath}
\usepackage{amssymb}
\usepackage{amsthm}
\usepackage{amsfonts}
\usepackage{bm}
\usepackage{xfrac}
\usepackage{empheq}
\usepackage[normalem]{ulem} 
\usepackage{soul} 
\usepackage{mathtools}

\usepackage{etoolbox}
\makeatletter
\patchcmd{\@makecaption}
{\scshape}
{}
{}
{}
\makeatletter
\patchcmd{\@makecaption}
{\\}
{.\ }
{}
{}
\makeatother

\newtheorem{example}{Example}

\theoremstyle{definition}

\usepackage{xcolor}
\usepackage{color}
\graphicspath{{figs/}}
\begin{document}
	\title{Extended Functional Representation Lemma: A Tool For Privacy, Semantic Representation, Caching, and Compression Design}
	\vspace{-5mm}
	\author{
		\IEEEauthorblockN{Amirreza Zamani, Mikael Skoglund \vspace*{0.5em}
			\IEEEauthorblockA{\\
				Division of Information Science and Engineering, KTH Royal Institute of Technology \\
				Email: \protect amizam@kth.se, skoglund@kth.se }}
	}
	\maketitle

	%
	\maketitle

\begin{abstract}
	This paper provides an overview of a problem in information-theoretic privacy mechanism design, addressing two scenarios in which private data is either observable or hidden. 
	In each scenario, different privacy measures are used, including bounded mutual information and two types of per-letter privacy constraints. 
	Considering the first scenario, 
	an agent observes useful data that is correlated with private data, and wants to disclose the useful information to a user.
	Due to the privacy concerns, direct disclosure is prohibited. Hence, a privacy mechanism is designed to generate disclosed data which maximizes the revealed information about the useful data while satisfying a privacy constraint. In the second scenario, the agent has additionally access to the private data. 
	We discuss how the Functional Representation Lemma, the Strong Functional Representation Lemma, and their extended versions are useful for designing low-complexity privacy mechanisms that achieve optimal privacy-utility trade-offs under certain constraints. Furthermore, another privacy design problem is presented where part of the private attribute is more private than the remaining part. Finally, we provide applications including semantic communications, caching and delivery, and compression designs, where the approach can be applied.  

\end{abstract}

\section{Introduction}
Information theory, a discipline born from the pioneering work of Claude Shannon in the 1940s, bridges the realms of engineering and mathematics. Its primary goal lies in quantifying, storing, and transmitting information effectively. Shannon's revolutionary insights formed the foundation for modern information theory, offering a rigorous mathematical framework to understand the capacities of communication systems and the fundamental limits of data compression and error correction. Consequently, this field finds practical applications in a wide range of applications, from telecommunications and data compression to privacy and machine learning, shaping the future of digital communication and information security.
The amount of data produced by networked sensors that record and analyze signals from physical environments, humans, robots, information processing and software systems is increasing rapidly. Disclosing unprocessed data may cause privacy risks  due to the adversarial inferences. 
Moreover, both theoretical and practical approaches to information-theoretic privacy (secrecy) can be utilized to protect information in various processing systems. Perfect privacy (secrecy) is often not attainable in applications, hence, we may need to relax the restriction \cite{shannon}. Altogether, for the disclosure of the data we need privacy mechanism designs.

 Considering privacy design problems through the lens of information theory, in \cite{makhdoumi}, the concept of a privacy funnel is introduced. 
 The privacy-utility trade-off, where both utility and privacy are measured using mutual information given a Markov chain, is studied in \cite{borz}. Under the assumption of perfect privacy, it is shown that the privacy mechanism design problem can be reduced to a linear program. 
 Moreover, in \cite{borz}, it has been shown that information can be only disclosed if the kernel (leakage matrix) between private data and useful data is not invertible. In \cite{khodam}, we generalize \cite{borz} by relaxing the perfect privacy assumption allowing some small bounded leakage. Specifically, we design privacy mechanisms with a per-letter (point-wise) privacy criterion considering an invertible kernel where a small bounded leakage is allowed. In \cite{Khodam22}, we generalized this result to a non-invertible kernel (leakage matrix). In \cite{kostala}, by using the Functional Representation Lemma bounds on privacy-utility trade-off for the two scenarios are derived. These results are obtained under the perfect privacy (secrecy) assumption, i.e., no leakages are allowed. When the private data is a deterministic function of the useful data, the bounds are tight.
  In \cite{shah}, the privacy problems studied in \cite{kostala} are extended by relaxing the perfect secrecy constraint and allowing some bounded leakage.

In this paper, we overview an approach that has been used in \cite{kostala} and \cite{shah} which is based on the Functional Representation Lemma (FRL), Strong Functional Representation Lemma (SFRL), Extended Functional Representation Lemma (EFRL) and Extended Strong Functional Representation Lemma (ESFRL).
We argue that since these lemmas have constructive proofs they can be useful for designing the privacy mechanisms which result in simple solutions to the high challenging problems.
 Moreover, we discuss how the approach can be used in a variety of applications including semantic communications, caching and delivery, fair and private representations of a database, and compression designs.
To do so we divide the overview into two main parts as follows\\
\textbf{Part I (\emph{Privacy-utility trade-off with non-zero leakage}):} First, we consider the problem studied in \cite{shah} which is closely related to \cite{kostala}, where the problem of \emph{secrecy by design} is studied. The privacy problems considered in \cite{kostala} are extended by relaxing the perfect privacy (secrecy) constraint and allowing some bounded leakage. More specifically, bounded mutual information has been considered for privacy leakage constraint. As argued in \cite{shah}, we can extend the Functional Representation Lemma and the Strong Functional Representation Lemma, introduced in \cite{kosnane}, by relaxing the independence condition to obtain lower bounds for the second scenario. 
It has been shown that under some assumptions the obtained privacy mechanism are optimal. 
 Furthermore, in the special case of perfect privacy (secrecy), the approach is discussed and compared with \cite{kostala}.\\ 
\textbf{Part II (\emph{Privacy-utility trade-off with non-zero leakage and per-letter privacy constraints}):}\\ In the second part, two different per-letter (point-wise) privacy constraints have been used in each scenario, replacing the bounded mutual information constraint. As discussed in \cite{Khodam22}, protecting private data at an individual level, rather than merely on average, can be more desirable. 
As argued in \cite{shah}, similar results as the extended versions of the Functional Representation Lemma and the Strong Functional Representation Lemma can be found in the previous part considering the per-letter privacy constraint rather than bounded mutual information. Using these results we can find lower bounds for the privacy-utility trade-off in the second scenario. Furthermore, bounds for three other problems are provided.\\ 
Moreover, we discuss a third problem called `\emph{Privacy-utility trade-off with non-zero leakage and prioritized private data}' which is studied in \cite{shah}. In this problem, considering the observable private data scenario, we assume that the private data is divided into two parts, where the first part is more private than the second part. Similar to the previous parts it has been shown that by using the extended versions of Functional Representation Lemma and the Strong Functional Representation Lemma we obtain lower bounds. Considering prioritized private data, the key idea to obtain lower bounds on the privacy-utility trade-off is to apply a randomization technique over part of the private data, rather than the private data.
Finally, we study applications where EFRL and ESFRL can be used. 

As an application, semantic communication involves transmitting a modified version of the original information source with reduced dimensionality to a receiver, whose goal is to extract information for a specific task \cite{sherdeniz}. 
Semantic communication goes beyond the literal message, taking into account its context, connotations, and nuances to minimize ambiguity and misunderstandings. It aligns the message with the recipient's perception, expectations, knowledge, and cultural context \cite{sherdeniz}. Recently, semantic communications have received significant research attention as a method to reduce the data load in 6G and future networks by transmitting only the semantically relevant information to the receiver.
As we mentioned earlier, another important dimension of emerging communication systems is data privacy. With increasing reliance on data for a wide range of applications, it is crucial for users to only reveal non-sensitive information to receivers. In this regard, semantic communication inherently addresses some privacy concerns by preventing the transmission of sensitive information. However, certain tasks may still require sending information correlated with private attributes, which can be challenging to identify and protect.
In \cite{zamani2024semantic}, we utilize concepts from the privacy mechanism design outlined in \cite{shah} 
to introduce an innovative private semantic communication framework. The proposed scheme offers a mathematical approach to design a goal-oriented private utility function. This function not only facilitates the receiver in achieving the goal but also guarantees the privacy of the data from the recipient.

Moreover, considering the compression problem with privacy constraints, a notion of perfect secrecy is introduced in \cite{shannon} by Shannon, where the public and private data are statistically independent. In the Shannon cipher system, one of $M$ messages is sent over a channel wiretapped by an eavesdropper, and it is shown that perfect secrecy is achievable if and only if the shared secret key length is at least $M$ \cite{shannon}. 
Moreover, in \cite{kostala}, the problems of variable length and fixed length compression have been studied and upper and lower bounds on the average length of encoded message have been derived. These results are derived under the assumption that the private data is independent of the encoded message, i.e., perfect secrecy. A similar approach has been used in \cite{kostala2}, where in a lossless compression problem the relations between secrecy, shared key and compression considering perfect secrecy, secrecy by design, mutual information leakage, maximal leakage and local differential privacy have been studied. 

Another application is caching and delivery design problems in cache-aided networks. One approach to reduce peak traffic and network congestion is to use distributed memories across the network \cite{maddah1}. Specifically, in \cite{maddah1}, a cache-aided network consisting of a single server connected
to several users through a shared bottleneck error-free link, is
considered, and the rate-memory trade-off has been characterized within a constant gap. 
In \cite{zamani2023cache}, we have generalized the 
problem considered in \cite{maddah1} by considering correlation between the private data and the database. To find an alternative solution to cache-aided data delivery problems in the presence of an adversary, we use
variable-length lossless compression techniques as in \cite{kostala}. The key idea is to use a two-part code construction, which is based on the Functional Representation Lemma (FRL) and one-time pad coding to hide the information about the private data and reconstruct the demanded files at user side.

\begin{figure}[]
	\centering
	\includegraphics[scale = .18]{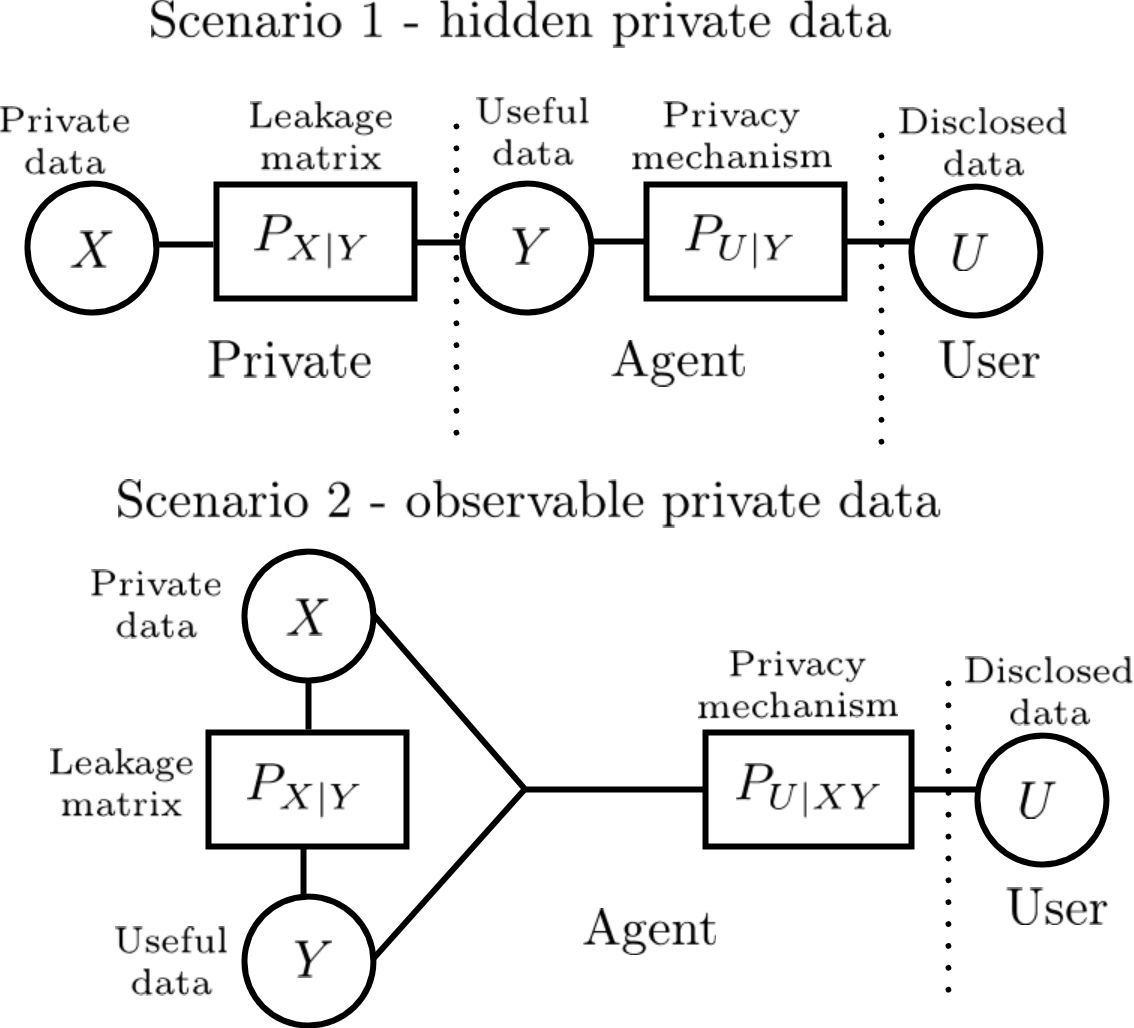}
	\caption{In the first scenario the agent has only access to $Y$ and in the second scenario the agent has access to both $X$ and $Y$.}
	\label{ISITsys}
\end{figure}
\section{System model} \label{sec:system}
In this paper, we present the problems that are studied in \cite{shah}, \cite{shahab}, \cite{borz}, \cite{7888175}, \cite{kostala}, and \cite{Khodam22}. Let random variable (RV) $Y$ denote the useful data which is correlated with the private data denoted by RV $X$. Furthermore, disclosed data is described by RV $U$. Two scenarios are considered. In both scenarios, as shown in Fig.~\ref{ISITsys}, an agent wants to disclose the useful information to a user. In the first scenario, the agent observes $Y$ and has no direct access to $X$, i.e., the private data is hidden. The goal is to design $U$ based on $Y$ that reveals as much information as possible about $Y$ and satisfies a bounded privacy criterion. In the second scenario, the agent has access to both $X$ and $Y$, i.e., the private data is observable, and can design $U$ based on $(X,Y)$ to release as much information as possible about $Y$ while satisfying the bounded leakage constraint.
In both scenarios we consider different privacy leakage constraints.
\subsection{Privacy-utility trade-off with non-zero leakage}
In this section, we consider the problems studied in \cite{shah}, \cite{shahab}, \cite{borz}, \cite{7888175} and \cite{kostala}. For both design problems we use mutual information as utility and leakage measures. The privacy mechanism design problems for the two scenarios can be stated as follows
\begin{align}
g_{\epsilon}(P_{XY})&=\sup_{\begin{array}{c} 
	\substack{P_{U|Y}:X-Y-U\\ \ I(U;X)\leq\epsilon,}
	\end{array}}I(Y;U),\label{main2}\\
h_{\epsilon}(P_{XY})&=\sup_{\begin{array}{c} 
	\substack{P_{U|Y,X}: I(U;X)\leq\epsilon,}
	\end{array}}I(Y;U).\label{main1}
\end{align} 
When the privacy mechanism has access to both the private data and the useful data, the function $h_{\epsilon}(P_{XY})$ is used. Moreover, the function $g_{\epsilon}(P_{XY})$ is used when the privacy mechanism has only access to the useful data. 
	When $\epsilon=0$, \eqref{main2} leads to the perfect privacy problem studied in \cite{borz}. It has been shown that for a non-invertible kernel $P_{X|Y}$, $g_0(P_{XY})$ can be obtained by a linear program.
	Additionally, for $\epsilon=0$, \eqref{main1} leads to the secret-dependent perfect privacy function $h_0(P_{XY})$, studied in \cite{kostala}, where upper and lower bounds on $h_0(P_{XY})$ have been derived.
	In \cite{shah}, we have extended the privacy problems considered in \cite{kostala} by relaxing the perfect secrecy constraint and allowing some leakage.
	 As mentioned earlier, \eqref{main2} and \eqref{main1} have been studied in previous works, e.g., \cite{borz}, and \cite{shahab}. Upper and lower bounds on $g_{\epsilon}(P_{XY})$ have been derived in \cite[Lemma 1]{shahab}, where we have
	 \begin{align}\label{coon}
	 \frac{H(Y)}{I(X;Y)}\epsilon\leq g_{\epsilon}(P_{XY})\leq H(Y|X)+\epsilon.
	 \end{align}
	 By using \cite[Lemma 2]{shahab} we have that the mapping $\epsilon\rightarrow g_{\epsilon}(P_{XY})$ is concave for any $\epsilon\geq 0$. Using the concavity of $g_{\epsilon}(P_{XY})$ leads to the lower bound in \eqref{coon}, since in this case, when $\epsilon=I(X;Y)$, $U=Y$ is feasible and the utility $H(Y)$ is achieved. Hence, the line $\frac{H(Y)}{I(X;Y)}\epsilon$ is attained. As stated in \cite[Remark 1]{shahab}, by using the concavity of $g_{\epsilon}(P_{XY})$, the lower bound \eqref{coon} can be improved. In this case, we have
	 \begin{align}
	 \epsilon\frac{H(Y)}{I(X;Y)}+g_0(P_{XY})\left(1-\frac{\epsilon}{I(X;Y)}\right)\leq g_{\epsilon}(P_{XY}).\label{coon2}
	 \end{align}
	 One benefit of the lower bound in \eqref{coon2} is that it can be used as a lower bound on $h_{\epsilon}(P_{XY})$, c.f., see \cite[Theorem 2]{shah}. Considering perfect privacy, i.e., $\epsilon=0$, $g_{0}(P_{XY})$ can be obtained by solving a linear program \cite[Theorem 1]{borz}. This means the optimal mapping is the solution to a linear program. Moreover, it has been shown that to find the optimal privacy-preserving mapping, it is sufficient to consider $U$ such that $|\mathcal{U}|\leq \text{null}(P_{X|Y})+1$. As the size of $\cal Y$ or $\cal X$ increases, solving the linear program proposed in \cite{borz} can become challenging. Therefore, simple upper and lower bounds are provided in \cite[Corollary 2]{borz} as follows
	 \begin{align*}
	 &(H(Y)-\log (\text{rank}(P_{X|Y})))^{+}\leq g_{0}(P_{XY})\leq \\&\min\{H(Y|X),\log(\text{null}(P_{X|Y})+1)\},
	 \end{align*}  
	 where $a^+=\begin{cases}
	 0, \ a<0,\\
	 a, \ a\geq 0
	 \end{cases}$. The upper bound $\log(\text{null}(P_{X|Y})+1)$ is obtained by using the sufficiency condition $|\mathcal{U}|\leq \text{null}(P_{X|Y})+1$, since in this case we have $H(U)\leq \log(\text{null}(P_{X|Y})+1)$. In \cite{calmon4}, necessary and sufficient conditions for attaining non-zero utility when the private data is hidden, i.e., $g_{0}(P_{XY})>0$, have been derived. It has been shown that $g_{0}(P_{XY})>0$ if and only if rows of $P_{X|Y}$ are linearly dependent. This result has been also shown and generalized in \cite{borz}, e.g., see \cite[Proposition 1]{borz}. Moreover, in \cite[Theorem 5]{kostala}, considering observable private data scenario, necessary and sufficient conditions for attaining non-zero utility, i.e., $h_{0}(P_{XY})>0$, have been derived. It has been shown that $h_{0}(P_{XY})>0$ if and only if $Y$ is not a deterministic function of $X$, i.e., $H(Y|X)>0$.
\subsection{Privacy-utility trade-off with non-zero leakage and per-letter privacy constraints}
In this part, we present the problems that are studied in \cite{shah} and \cite{Khodam22}. The privacy mechanism design problems for the two scenarios are stated as follows 
\begin{align}
g_{\epsilon}^{w\ell}(P_{XY})&=\sup_{\begin{array}{c} 
	\substack{P_{U|Y}:X-Y-U\\ \ d(P_{X,U}(\cdot,u),P_XP_{U}(u))\leq\epsilon,\ \forall u}
	\end{array}}I(Y;U),\label{main22}\\
h_{\epsilon}^{w\ell}(P_{XY})&=\sup_{\begin{array}{c} 
	\substack{P_{U|Y,X}: d(P_{X,U}(\cdot,u),P_XP_{U}(u))\leq\epsilon,\ \forall u}
	\end{array}}I(Y;U),\label{main11}\\
g_{\epsilon}^{\ell}(P_{XY})&=\sup_{\begin{array}{c} 
	\substack{P_{U|Y}:X-Y-U\\ \ d(P_{X|U}(\cdot|u),P_X)\leq\epsilon,\ \forall u}
	\end{array}}I(Y;U),\label{main222}\\
h_{\epsilon}^{\ell}(P_{XY})&=\sup_{\begin{array}{c} 
	\substack{P_{U|Y,X}: d(P_{X|U}(\cdot|u),P_X)\leq\epsilon,\ \forall u}
	\end{array}}I(Y;U),\label{main12}
\end{align} 
where $d(P,Q)$ corresponds to the total variation distance between two distributions $P$ and $Q$, i.e., $d(P,Q)=\sum_x |P(x)-Q(x)|$.
We can use the functions $h_{\epsilon}^{w\ell}(P_{XY})$ and $h_{\epsilon}^{\ell}(P_{XY})$ when the privacy mechanism has access to both the private data and the useful data. Moreover, we use the functions $g_{\epsilon}^{w\ell}(P_{XY})$ and $g_{\epsilon}^{\ell}(P_{XY})$ when the privacy mechanism has only access to the useful data. As we outlined in \cite{shah}, the point-wise privacy constraints used in \eqref{main22} and \eqref{main222}, i.e., $d(P_{X,U}(\cdot,u),P_XP_U(u))\leq\epsilon,\ \forall u,$ and $d(P_{X|U}(\cdot|u),P_X)\leq\epsilon,\ \forall u,$ are called the \emph{weighted strong $\ell_1$-privacy criterion} and the \emph{strong $\ell_1$-privacy criterion}. We refer to them as strong since they are per-letter (point-wise) privacy constraints, i.e., they must hold for every $u\in\cal U$. The difference between the two privacy constraints in this work is the weight $P_U(u)$, therefore, we refer to $d(P_{X,U}(\cdot,u),P_XP_U(u))\leq\epsilon,\ \forall u,$ as weighted.
	In \cite{Khodam22}, we have used the leakage constraint $d(P_{X|U}(\cdot|u),P_X)\leq\epsilon,\ \forall u$, where we called it the \emph{strong $\ell_1$-privacy criterion}.  
	When $\epsilon=0$, both \eqref{main22} and \eqref{main222} lead to the perfect privacy problem studied in \cite{borz}. 
	For $\epsilon=0$, both \eqref{main11} and \eqref{main12} lead to the secret-dependent perfect privacy function $h_0(P_{XY})$, studied in \cite{kostala}, where upper and lower bounds on $h_0(P_{XY})$ have been derived. In \cite{shah}, we have improved these bounds.
	The privacy problem defined in \eqref{main222} has been studied in \cite{Khodam22} where we have provided a lower bound on $g_{\epsilon}^{\ell}(P_{XY})$ using the information geometry concepts. 
	Next, we discuss the motivation for selecting point-wise privacy leakage measures.

As we outlined in \cite{shah} and \cite{Khodam22}, for small $\epsilon$, both privacy constraints mean that $X$ and $U$ are almost independent. As we discussed in \cite{Khodam22}, closeness of $P_{X|U}(\cdot|u)$ and $P_X$ allows us to approximate $g_{\epsilon}^{\ell}(P_{XY})$ with a Taylor series expansion and find a lower bound. In \cite{shah}, we have shown that by using a similar methodology, we can approximate $g_{\epsilon}^{w\ell}(P_{XY})$ exploiting the closeness of $P_{X,U}(\cdot,u)$ and $P_XP_U(u)$. This provides us a lower bound for $g_{\epsilon}^{w\ell}(P_{XY})$. 
Next, we provide a list of properties of the weighted strong $\ell_1$-privacy criterion and the strong $\ell_1$-privacy criterion.
\begin{itemize}
	\item The strong $\ell_1$-privacy criterion and the weighted strong $\ell_1$-privacy criterion satisfy the linkage inequality \cite{shah}.
	\item The average of the weighted strong $\ell_1$-privacy criterion and the strong $\ell_1$-privacy criterion with weights equal one and $P_U(u)$, respectively, satisfy the post-processing inequality \cite{shah}.
	\item The weighted strong $\ell_1$-privacy criterion and the strong $\ell_1$-privacy criterion result in bounded adversarial inference performance that is modeled in \cite{Calmon1}. 
	\item Another property is the relation between the strong $\ell_1$-privacy criterion and $\text{MMSE}(X|U)$ which has been used in \cite[Corrolary~2]{MMSE}. An interesting result is that the strong $\ell_1$-privacy criterion yields a lower bound on $\text{MMSE}(X|U)$ \cite[Proposition 9]{Khodam22}.
\end{itemize}
As outlined in \cite[page 4]{Total} and \cite{shah}, one advantage of the linkage inequality is to keep the privacy in layers of private information which is discussed in the following. Consider the scenario where the Markov chain $X-Y-U$ holds, but the distribution of $X$ is unknown. If we can find $\tilde{X}$ such that $X-\tilde{X}-Y-U$ holds and the distribution of $\tilde{X}$ is known then by using the linkage inequality we conclude $\mathcal{L}(X;U=u)\leq \mathcal{L}(\tilde{X};U=u)$, where $\mathcal{L}(A,B=b)$ denotes the point-wise leakage from $A$ to $b$. In other words, if the framework is designed for $\tilde{X}$, then a privacy constraint on $\tilde{X}$ yields the constraint on $X$, hence, provides an upper bound for any pre-processed RV $X$. To have the Markov chain $X-\tilde{X}-Y-U$ let $\tilde{X}$ be the private data and $X$ be a function of private data which is not known. For instance, let $\tilde{X}=(X_1,X_2,X_3)$ and $X=X_1$. Thus, the mechanism that is designed based on $\tilde{X}-Y-U$ preserves the privacy leakage constraint on $X$ and $U$. As outlined in \cite[Remark 2]{Total}, among all the $L^p$-norms ($p\geq 1$), only the $\ell_1$-norm satisfies the linkage inequality.
 Practical examples based on the MNIST dataset and a medical experiment with real data are provided in \cite[Experiment 2]{Khodam22} and \cite[Experiment 3]{Khodam22}, demonstrating the applicability of the strong $\ell_1$-privacy criterion. Meaningful interpretations of the solution to $g_{\epsilon}^{\ell}(P_{XY})$ are discussed for both experiments in \cite{Khodam22}. Additionally, a geometrical interpretation of the solution has been presented in \cite[Section IV]{Khodam22}. Notably, all candidate optimizers for $g_{\epsilon}^{\ell}(P_{XY})$ lie within an $\ell_1$-ball with a bounded radius and specific centers \cite[Section IV]{Khodam22}. The role of the strong $\ell_1$-privacy criterion can be evaluated using various metrics, such as the probability of error and MMSE, for measuring both utility and privacy leakage in $g_{\epsilon}^{\ell}(P_{XY})$, and the results can be compared with previous works. Further details are available in \cite[Section V-B]{Khodam22}.

As we outlined in \cite{shah}, another property of the $\ell_1$-distance is the relation between the probability of error and the $\ell_1$-norm in a hypothesis test. For the binary hypothesis test with $H_0:X\sim P$ and $H_1:X\sim Q$,
the expression $1-TV(P, Q)$ is the sum of missed detection and false alarm probabilities. Thus, we have
$
TV(P,Q)=1-2P_e,
$
 where $P_e$ is the probability of error (the probability that we can not decide the right distribution for $X$ with equal prior probabilities for $H_0$ and $H_1$). For instance, consider a scenario where we want to determine whether $X$ and $U$ are independent or correlated. To this end, let $P=P_{X,U}$, $Q=P_{X}P_{U}$, with hypotheses $H_0:X,U\sim P$ and $H_1:X,U\sim Q$. We have
\begin{align*}
TV(P_{X,U};P_{X}P_{U})&=\frac{1}{2}\sum_u P_U(u)\left\lVert P_{X|U=u}-P_X \right\rVert_1\!\\&\leq \frac{1}{2}\epsilon.
\end{align*}
Thus, as the leakage increases, meaning that $TV(P_{X,U};P_{X}P_{U})$ also increases, the probability of error decreases.  \\
Finally, when using $\ell_1$-distance as the measure of privacy leakage, after approximating $g_{\epsilon}^{w\ell}(P_{XY})$ and $g_{\epsilon}^{\ell}(P_{XY})$, the resulting problems reduce to linear programming, which are significantly easier to solve.
\subsection{Privacy-utility trade-off with non-zero leakage and prioritized private data}
In this part, we consider the problem is \cite{shah}[Section II-C]. 
The private data $X$ is divided into two parts $X_1$ and $X_2$, where the first part is more private than the other part, i.e., the privacy leakage of $X_1$ is less than or equal to the privacy leakage of $X_2$. Mutual information for measuring both utility and privacy leakage is used and we only consider the second scenario where the private data $X$ is observable. Hence, the problem can be stated as follows
   \begin{align}
   h_{\epsilon}^{p}(P_{X_1X_2Y})&=\sup_{\begin{array}{c} 
   	\substack{P_{U|YX_1X_2}: I(U;X_1,X_2)\leq\epsilon,\\ I(U;X_1)\leq I(U;X_2)}
   	\end{array}}I(Y;U).\label{main111}
   \end{align}
The constraint $I(U;X_1,X_2)\leq\epsilon$ ensures that the total leakage remains bounded by $\epsilon$, while the constraint $I(U;X_1)\leq I(U;X_2)$ reflects the prioritization of $X_1$. In practice, private data often has varying levels of privacy leakage, and in this case, we consider two distinct levels.
	For $\epsilon=0$, \eqref{main111} leads to the secret-dependent perfect privacy function $h_0(P_{XY})$.
\section{Relation between the problems}\label{rel}
The relation between the problems outlined earlier has been discussed in \cite{shah}[Section III]. We first have obtained the relation between the strong $\ell_1$-privacy criterion, the weighted strong $\ell_1$-privacy criterion and bounded mutual information. For instance, Pinsker's inequality \cite{verdu} can be used to derive an inequality between the weighted strong $\ell_1$-privacy measure and mutual information. The relations can help us to find an alternative approach to solve each problem. For example, if the problem with bounded mutual information is hard to address, one can apply an information geometry approach to address the strong $\ell_1$-privacy criterion, which in turn provides a bound on the problem with bounded mutual information.
\section{Main approach to design the privacy mechanisms}\label{result}
In this section, we discuss how EFRL \cite[Lemma 4]{shah} and ESFRL \cite[Lemma 5]{shah} help us to obtain simple privacy designs for the problems outlined earlier. 
As discussed in \cite{shah}, the idea of extending Functional Representation Lemma and Strong Functional Representation Lemma is basically adding a randomized response argument to the random variable $U$ found by the FRL and the SFRL. The idea is simple, when it can be connected with an old principle it gets creditable.
As an example we present two simple bounds on $h_{\epsilon}(P_{XY})$ derived in \cite[Theorem 2]{shah} which corresponds to the problem \emph{privacy-utility trade-off with non-zero leakage}. Before obtaining the bounds we derive an expression for $I(Y;U)$. We have
\begin{align}
I(Y;U)&=I(X;U)+H(Y|X)\nonumber\\&-H(Y|U,X)-I(X;U|Y).\label{key5}
\end{align}
As argued in \cite{kostala} and \cite{shah}, \eqref{key5} is an important observation to find lower and upper bounds for $h_{\epsilon}(P_{XY})$ and $g_{\epsilon}(P_{XY})$.
The equation \eqref{key5} can be used to find a simple upper bound as follows. Since the leakage is bounded by $\epsilon$ we have
\begin{align*}
I(Y;U)\leq \epsilon+H(Y|X).
\end{align*}
To find the first lower bound let $U^*$ be the RV found by the EFRL , then
\begin{align*}
I(U^*;X)=\epsilon,\\
H(Y|U^*,X)=0.
\end{align*}
Then by using \eqref{key5} we have
\begin{align*}
I(Y;U^*)&=\epsilon+H(Y|X)-I(X;U^*|Y)\\ &\geq \epsilon+H(Y|X)-H(X|Y).
\end{align*}
Furthermore, we can find a second lower bound by using ESFRL instead of EFRL. To do so, let $\bar{U}$ be the RV found by the EFRL , then
\begin{align*}
I(\bar{U};X)&=\epsilon,\\
H(Y|\bar{U},X)&=0,\\
I(X;\!\bar{U}|Y)&\!\leq\! \alpha H(X|Y)\!+\!(1\!-\!\alpha)\!\left[ \log(I(X;\!Y)\!+\!1)\!+\!4\right],
\end{align*}
where $\alpha=\frac{\epsilon}{H(X)}$ is the probability of randomization over the private data $X$.
By using \eqref{key5} we have
 \begin{align*}
 I(Y;\bar{U})&=\epsilon+H(Y|X)-I(X;\bar{U}|Y)\\ &\geq \epsilon+H(Y|X)-\alpha H(X|Y)\\&-(1-\alpha)\left( \log(I(X;Y)+1)+4 \right).
 \end{align*}
It can be seen that in the second lower bound, the impact of the term $H(X|Y)$ is reduced; however, the penalty is $(1-\alpha)\left( \log(I(X;Y)+1)+4 \right)$. As argued in \cite[Example 1]{shah}, when $X$ and $Y$ are independent and $H(X)>4$, the second lower bound is dominant. Furthermore, when $X$ is a deterministic function of $Y$, the first lower bound is dominant \cite[Example 2]{shah}. Regarding the tightness of the lower bounds, it has been shown in \cite[Theorem 2]{shah} that the first lower bound is tight when $H(X|Y)=0$, i.e., when $X$ is a deterministic function of $Y$. Furthermore, if the first lower bound is tight, it follows that $H(X|Y)=0$. In other words, when $X$ is deterministic function of $Y$, the first lower bound attains the optimal privacy-utility trade-off. As discussed in \cite{shah}, when non-zero leakage is allowed, the EFRL and the ESFRL can significantly improve the utility compared with the FRL and the SFRL, respectively. Using \eqref{key5} utility attained by the FRL is $H(Y|X)-H(X|Y)$, which is less than or equal to utility attained by EFRL. Furthermore, the utility achieved by the SFRL is $H(Y|X)-\left( \log(I(X;Y)+1)+4 \right)$, which is less than or equal to utility attained by ESFRL. Next, we discuss a special case where $\epsilon=0$. In this case, by letting $\epsilon=0$ in the outlined bounds we obtain two lower bounds where the first lower bound has been derived in \cite[Theorem 6]{kostala} and the second lower bound is proposed by \cite{shah}. Note that the bound proposed by \cite{shah} has generalized the results in \cite{kostala}. Furthermore, in \cite{shah}, we have obtained a new upper bound by using the definition of \emph{excess functional information} in
\cite{kosnane}. Considering the case where $X$ is a deterministic function of $Y$, it is shown that both upper bounds attain the same utilities. Furthermore, it has been shown that when the useful data is a binary RV the new bound is attained. To see a benefit of this result, we recall an example from \cite{shah}. As shown in \cite{shah}, the new upper bound proposed by \cite{shah} can improve \cite{kostala}.
\begin{figure}[]
	\centering
	\includegraphics[scale = .08]{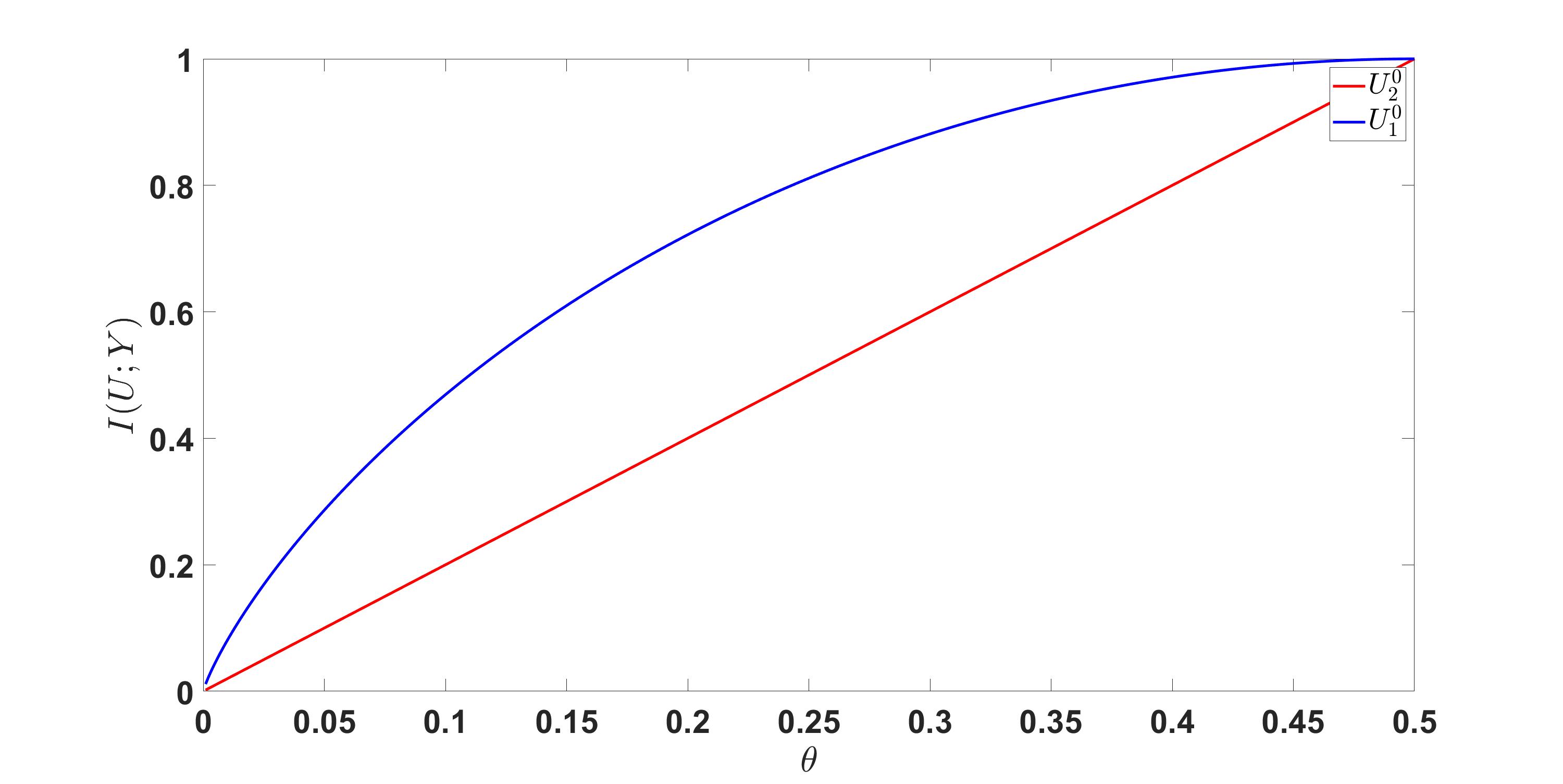}
	\caption{Comparing the upper bounds obtained in \cite{kostala} and \cite{shah} for $BSC(\theta)$. The blue curve illustrates the upper bound found in \cite{kostala} and the red line shows the upper bound found in \cite{shah}.}
	\label{fig:kir}
\end{figure}
\begin{example}(\cite[Example 3]{shah})
	Let the binary RVs $X\in\{0,1\}$ and $Y\in\{0,1\}$ have the following joint probability distribution
	\begin{align*}
	P_{XY}(x,y)=\begin{cases}
	\frac{1-\theta}{2}, \ &x=y\\
	\frac{\theta}{2}, \ &x\neq y
	\end{cases},
	\end{align*}
	where $\theta<\frac{1}{2}$. In this case, the red curve attains the optimal trade-off and for constructing the RV that achieves the bound we use Poisson functional representation \cite{kosnane}. 
\end{example}
Next, we discuss more on tightness of the previous upper bound on $h_{\epsilon}(P_{XY})$. Earlier, we mentioned that when $X$ is deterministic function of $Y$, the optimal privacy-utility trade-off can be attained. As we outlined in \cite{shah}, we can find a larger set of distributions $P_{XY}$ compared to the constraint `$X$ is deterministic function of $Y$', for which the optimal trade-off is achieved. To do so, we have used the concept of \emph{common information}, where both notions of the Wyner \cite{wyner} or G{\'a}cs-K{\"o}rner \cite{gacs1973common} common information can be used. We have shown that when the common information and the mutual information between $X$ and $Y$ are equal, the upper bound on $h_{\epsilon}(P_{XY})$ can be achieved and we have $h_{\epsilon}(P_{XY})=\epsilon+H(Y|X)$. Noting that if $X$ is a deterministic function of $Y$, then the common information and mutual information between $X$ and $Y$ are equal. Hence, we have obtained a more general condition to find the optimal trade-off. Consequently, we studied a case where $Y$ is a deterministic function of $X$. We have shown that in this case $h_{\epsilon}(P_{XY})=\epsilon$ and proposed a privacy mechanism design which attains the optimal solution. The proposed scheme is based on using randomization over the useful data $Y$. 
An interesting result is that considering zero leakage scenario, $\epsilon=0$, the constraint $H(Y|X)=0$ results in zero utility, i.e., $h_0(P_{XY})=0$, \cite[Theorem 5]{kostala}. This result can be verified by using \eqref{key5}. However, when non-zero leakage is allowed we can achieve non-zero utility. As argued in \cite{kostala} when $X$ is a deterministic function of $Y$, the necessary and sufficient conditions for achieving the optimal trade-off are fulfilled. Furthermore, this result holds when $X$ and $Y$ are independent or $Y$ is a deterministic function of $X$. However, in \cite{shah} we have shown that for any $0\leq \epsilon < I(X;Y)$, this statement can be generalized. The condition that $X$ is a deterministic function of $Y$ can be substituted by the condition that the common information and mutual information between $X$ and $Y$ are equal.
In \cite[Theorem 3]{shah}, we have shown the following equivalencies
\begin{itemize}
	\item [i.] $g_{\epsilon}(P_{XY})=H(Y|X)+\epsilon$,
	\item [ii.] $g_{\epsilon}(P_{XY})=h_{\epsilon}(P_{XY})$,
	\item [iii.] $h_{\epsilon}(P_{XY})=H(Y|X)+\epsilon$.
\end{itemize}
This implies that when $h_{\epsilon}(P_{XY})$ achieves the optimal trade-off, $g_{\epsilon}(P_{XY})$ also attains it, and vice versa. Consequently, when the common information and mutual information between $X$ and $Y$ are equal we have $g_{\epsilon}(P_{XY})=H(Y|X)+\epsilon$. In other words, when the common information and mutual information between $X$ and $Y$ are equal we have  
\begin{align*}
g_{\epsilon}(P_{XY})=h_{\epsilon}(P_{XY})=H(Y|X)+\epsilon.
\end{align*}
\begin{figure}[]
	\centering
	\includegraphics[width = 0.17\textwidth]{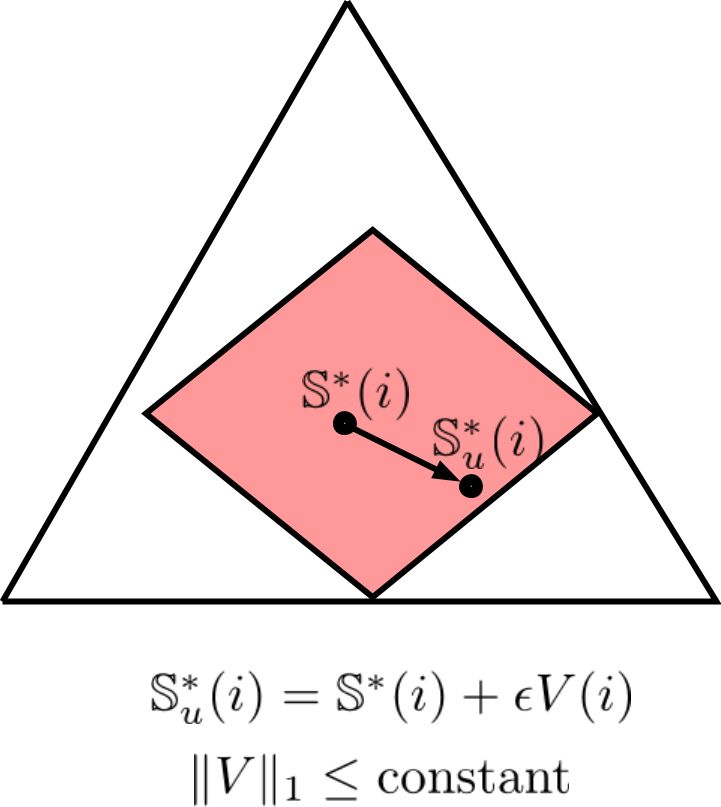}
	\caption{Possible positions of the optimizers. $\mathbb{S}_u^*(i)$ is inside an $\ell_1$-ball of radius $r$ with center $\mathbb{S}^*(i)$.}
	\label{pos}
\end{figure}
\subsection{Privacy-utility trade-off with non-zero leakage and per-letter privacy constraints}
In this section, we review the approaches which are used to address $g_{\epsilon}^{w\ell}(P_{XY})$, $h_{\epsilon}^{w\ell}(P_{XY})$, $g_{\epsilon}^{\ell}(P_{XY})$ and $h_{\epsilon}^{\ell}(P_{XY})$. As argued in \cite{shah}, to find lower bounds on $h_{\epsilon}^{w\ell}(P_{XY})$ we can extend the EFRL and the ESFRL by replacing the mutual information constraint, i.e., $I(U;X)=\epsilon$, with the weighted $\ell_1$-strong privacy criterion defined in \eqref{main22} and \eqref{main11}. Following this approach lead to the privacy designs for $h_{\epsilon}^{w\ell}(P_{XY})$. In \cite{Khodam22}, we have shown that $g_{\epsilon}^{\ell}(P_{XY})$ can be approximated by a linear program using information geometry concepts. Using this result we can derive a lower bound for $g_{\epsilon}^{\ell}(P_{XY})$. As illustrated in Fig. \ref{pos}, it has been shown that the optimizers of the linear program are inside an $\ell_1$-ball with bounded radius and fixed centers \cite{Khodam22}. As we discussed in \cite{shah}, by following a similar approach, i.e., using concepts from information geometry and approximating the utility, we can approximate $g_{\epsilon}^{w\ell}(P_{XY})$ by a linear program which leads to a privacy mechanism design. Moreover, in \cite{shah}, we have obtained upper bounds on the error of approximations. These error bounds lead to the upper bounds on $g_{\epsilon}^{\ell}(P_{XY})$. To find upper bounds on $h_{\epsilon}^{\ell}(P_{XY})$ and $h_{\epsilon}^{w\ell}(P_{XY})$ we use Pinsker and reverse Pinsker's inequalities \cite{verdu}.  

\begin{figure}[]
	\centering
	\includegraphics[width = .5\textwidth]{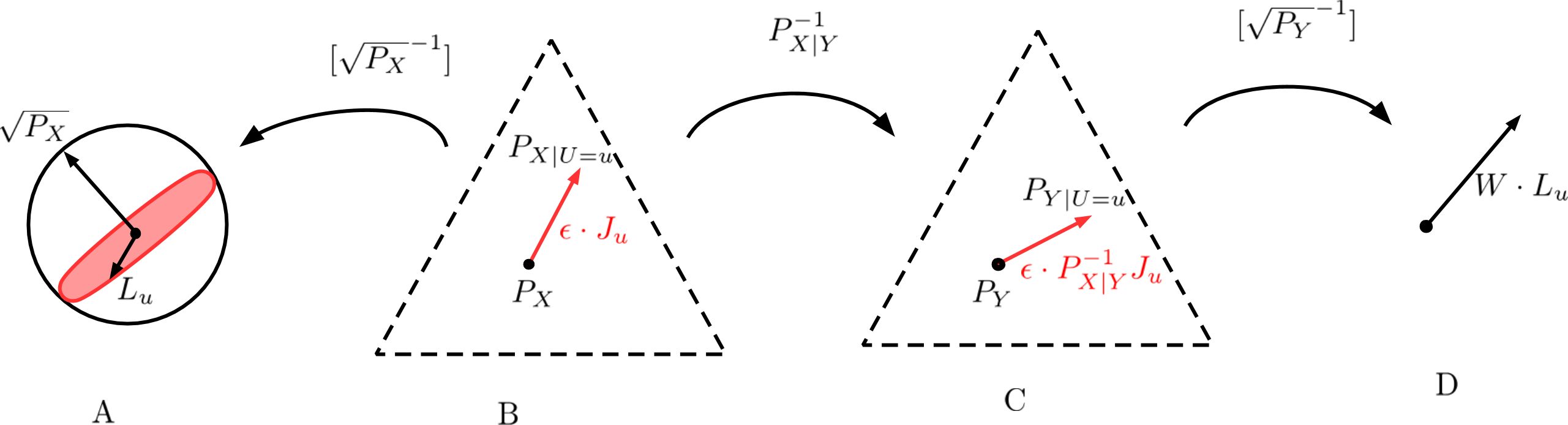}
	\caption{ For the privacy mechanism design, we are looking for $L^*$ in the red region (vector space A) which results in a vector with the largest Euclidean norm in vector space D. Space B and space C are probability spaces for the input and output distributions, the circle in space A represents the vectors that satisfy the strong $\chi^2$-privacy criterion and the red region denotes all vectors that are orthogonal to vector $\sqrt{P_X}$. Starting from Space A and reaching Space D the mapping between Space A and Space D can be found as  $W=[\sqrt{P_Y}^{-1}]P_{X|Y}^{-1}[\sqrt{P_X}]$.}
	\label{fig:inter}
\end{figure}

Replacing the $\ell_1$-norm with the $\chi^2$-distance in $g_{\epsilon}^{\ell}(P_{XY})$ leads to the privacy problem studied in \cite{khodam}. It has been shown that when the leakage matrix $P_{X|Y}$ is invertible the main problem can be approximated by a linear problem of finding maximum singular value of a matrix with corresponding singular vector. As shown in Fig. \ref{fig:inter}, the optimizers of the approximated problem must lie within an $\ell_2$-ball of radius 1, which are orthogonal to a fixed vector.    
\subsection{Privacy-utility trade-off with non-zero leakage and prioritized private data}
In this section we review the privacy designs obtained in \cite{shah}, that finds lower bounds on $h_{\epsilon}^{p}(P_{X_1X_2Y})$ defined in \eqref{main111}. We have discussed that by using similar approaches used in EFRL and ESFRL lower bounds can be obtained. The main idea is to add randomized response over $X_2$ instead of $X$. As we outlined in \cite{shah}, lower bounds on $h_{\epsilon}^{p}(P_{X_1X_2Y})$ can be used as lower bounds on $h_{\epsilon}^{p}(P_{X_1X_2Y})$ since we have $h_{\epsilon}^{p}(P_{X_1X_2Y})\leq h_{\epsilon}(P_{X_1X_2Y})$.
\section{Applications}
In this section, we provide a list of applications where the proposed approach in \cite{shah} can be applied.
\begin{itemize}
	\item Caching and delivery design with privacy constraints
	\item Multi-user privacy design with privacy constraints
	\item Compression design problems with privacy constraints
	\item Semantic communications with privacy constraints
	\item Multi-task semantic communications with privacy constraints
	\item Fair and private representations
\end{itemize}
\vspace{-5mm}
\subsection{Caching and delivery design with privacy constraints}
\begin{figure}[]
 	\centering
 	\includegraphics[scale = .11]{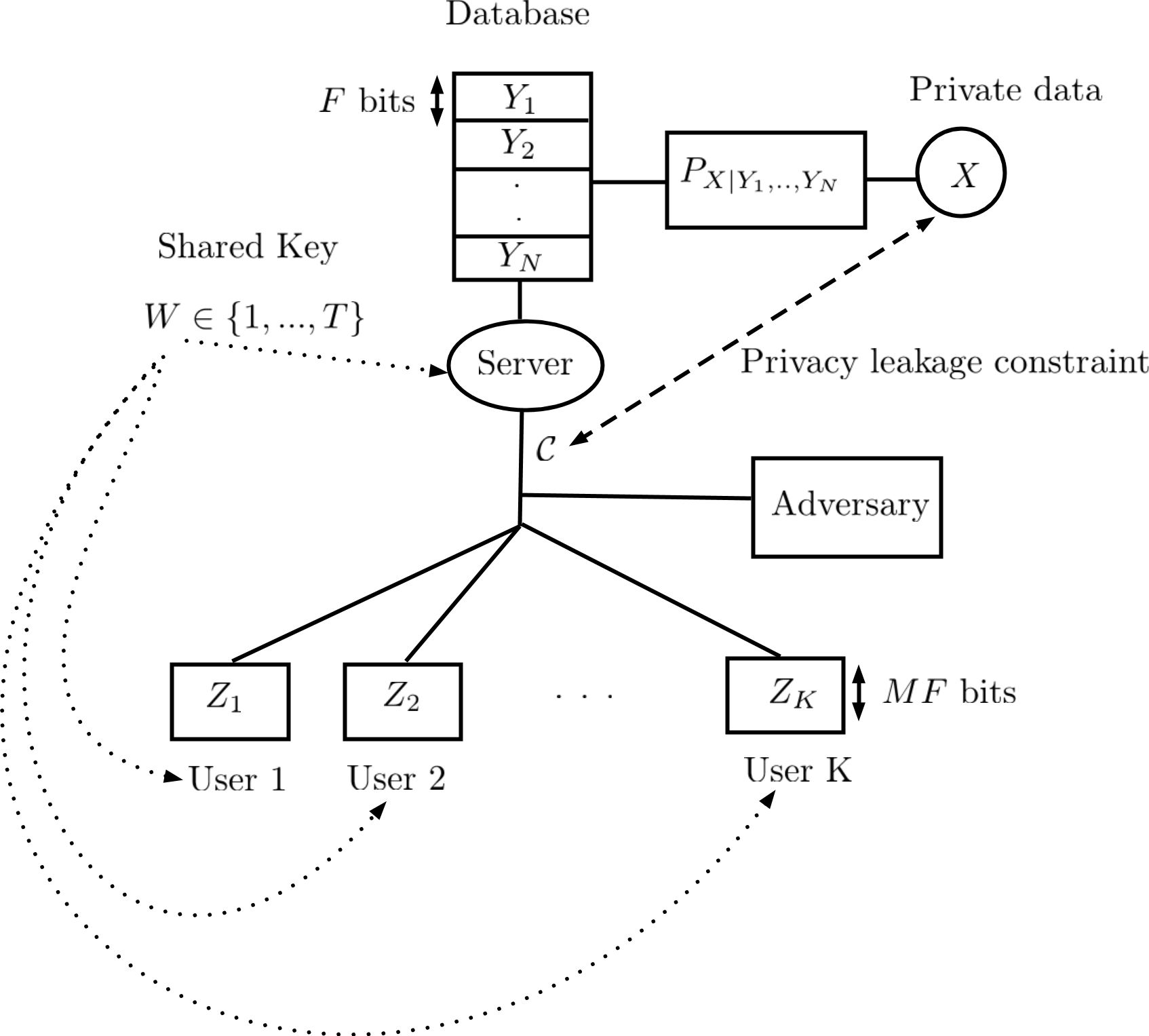}
 	\caption{A server wants to send a response over a shared link to satisfy users$'$ demands, but since the database is correlated with the private data existing schemes are not applicable. In the delivery phase, we hide the information about $X$ using one-time-pad coding and send the rest of response using Functional Representation Lemma (FRL).}
 	\label{wii}
 \end{figure}

In this section, we review the problem studied in \cite{zamani2023cache}, which is illustrated in Fig. \ref{wii}. A server has access to a database consisting of $N$ files $Y_1,..,Y_N$, where each file, of size $F$ bits, is sampled from the joint distribution $P_{XY_1\cdot Y_N}$. Random variable $X$ denotes the private data. We assume that the server knows the realization of the private variable $X$ as well. The server is connected to $K$ users through a shared link, where User $i$ has access to a local memory of size $MF$ bits. Furthermore, we assume that the server and the users have access to a shared secret key denoted by RV $W$, of size $T$. The system works in two phases: the placement and delivery phases, respectively, \cite{maddah1}. In the placement phase, the server fills the local caches using the database. Let $Z_k$ denote the content of the local cache memory of user $k$, $k\in[K]\triangleq\{1,..,K\}$ after the placement phase. In the delivery phase, first the users send their demands to the server, where $d_k\in[N]$ denotes the demand of user $k$. The server sends a response, denoted by $\mathcal{C}$, over the shared link to satisfy all the demands, simultaneously. We assume that an adversary has access to the shared link as well, and uses $\mathcal {C}$ to extract information about $X$. However, the adversary does not have access to the local cache contents or the secret key. 
Since the files in the database are all correlated with the private data $X$, the coded caching and delivery techniques proposed by \cite{maddah1} do not satisfy the privacy requirement.
 
The goal of the cache-aided private delivery problem is to find a response $\mathcal{C}$ with minimum possible average length which satisfies a certain privacy constraint and the zero-error decodability constraint of users. Here, the worst case demand combinations $d=(d_1,..,d_K)$ is considered to construct $\mathcal {C}$. The expectation is taken over the randomness in the database. In this study, first a perfect privacy constraint is considered, i.e., we require $\mathcal {C}$ to be independent of $X$. Let $\hat{Y}_{d_k}$ denote the decoded message of user $k$ using $W$, $\mathcal {C}$, and $Z_k$. User $k$ should be able to recover $Y_{d_k}$ reliably, i.e., $\mathbb{P}{\{\hat{Y}_{d_k}\neq Y_{d_k}\}}=0$, $\forall k\in[K]$. We remark that we have a cache-aided variable-length compression problem with a privacy constraint. Hence, to solve this problem  
techniques used in privacy mechanisms, data compression, and cache design and coded delivery problems have been utilized and they have been combined to build such a code. 
In particular, data compression techniques employed in \cite{kostala} and caching design techniques in \cite{maddah1} have been used. Finally, the results are extended for correlated $X$ and $\mathcal{C}$, i.e., when non-zero leakage is allowed. 
\vspace{-5mm}
\subsection{Multi-user privacy design with privacy constraints}
In this part, we consider the problem studied in \cite{multiAmir}. The problem is closely related to \cite{9457633}, where fundamental limits of private data disclosure are studied. In \cite{9457633}, the goal is to minimize the leakage under the utility constraints with non-specific tasks. It has been shown that under the assumption that the private data is an element-wise deterministic function of useful data, 
the main problem can be reduced to multiple privacy funnel (PF) problems. Moreover, the exact solution to each problem has been obtained.  \\
\begin{figure}[]
	\centering
	\includegraphics[scale = .12]{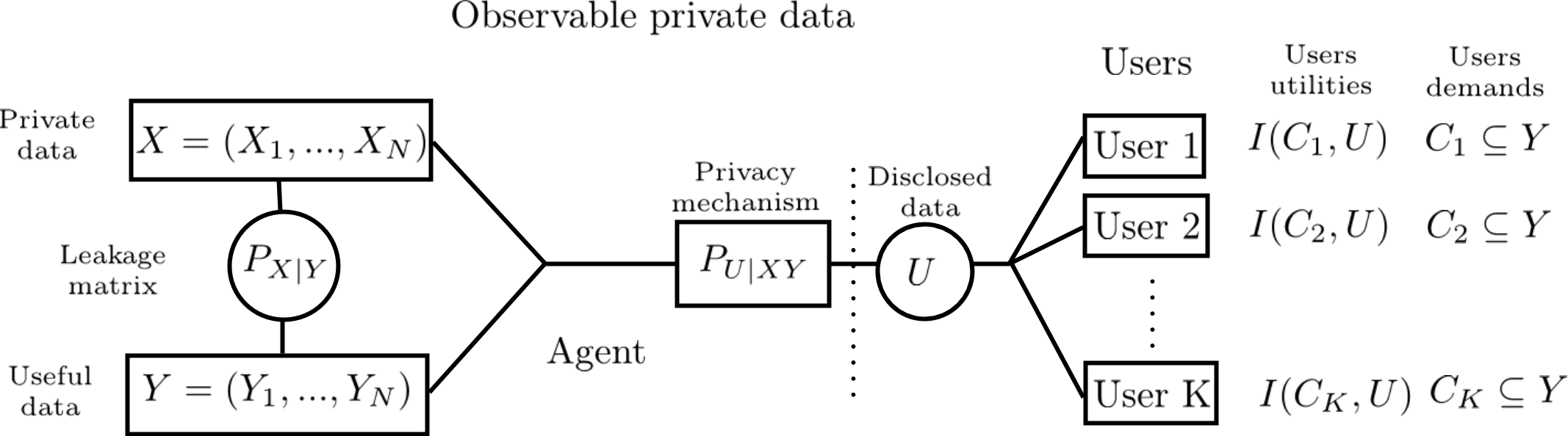}
	\caption{In this work the agent has access to $Y$ and $X$. Each user demands a sub-vector of $Y$ and the agent releases message $U$ which maximizes a linear combination of utilities.}
	\label{ITWsys}
\end{figure}
In this problem, $Y=(Y_1,...,Y_N)$ denotes the useful data where $Y_1,...,Y_N$ are mutually independent random variables. The useful data is correlated with the private data denoted by $X=(X_1,...,X_N)$ where $X_1,...,X_N$ are mutually independent RVs. As shown in Fig.~\ref{ITWsys}, User $i$, $i\in\{1,...,K\}$, demands an arbitrary sub-vector of $Y$ denoted by $C_i$ and an agent wishes to design disclosed data denoted by $U$ which maximizes a linear combination of utilities (weighted sum-utility) while satisfying the bounded leakage constraint, i.e., $I(X;U)\leq \epsilon$. Utility of user $i$ is measured by the mutual information between $C_i$ and $U$, i.e., $I(C_i,U)$. The problem is motivated by the dual of the privacy design problem studied in \cite{9457633}. The assumption that the private data is a deterministic function of useful data is very restrictive, hence, we generalize the assumption by letting $X_i$ and $Y_i$ be arbitrarily correlated.     
First, upper bounds on the privacy-utility trade-off are derived by using the same transformation as used in \cite[Theorem 1]{9457633}. Then, it is shown that the upper bound can be decomposed into $N$ parallel privacy design problems. Moreover, lower bounds on privacy-utility trade-off are obtained by using the Functional Representation Lemma and the Strong Functional Representation Lemma. The lower bound leads to a privacy mechanism design which is based on the randomization used in \cite{shah}. In \cite{shah}, we used randomization to extend FRL and SFRL relaxing the independence condition. 
It is shown that the upper bound is tight within a constant term and the lower bound is optimal when the private data is an element-wise deterministic function of the useful data, i.e., the mechanism design is optimal in this case. 
\subsection{Compression design problems with privacy constraints}
\begin{figure}[]
	\centering
	\includegraphics[scale = .55]{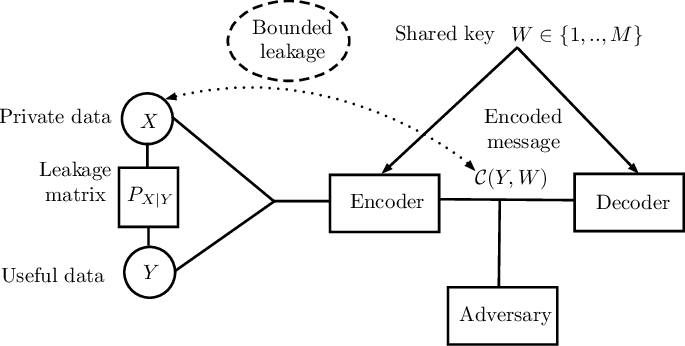}
	\caption{An encoder wants to compress $Y$ which is correlated with $X$ under certain privacy leakage constraints and sends it over a channel where an eavesdropper has access to the output of the encoder.} 
	\label{ITWsys1}
\end{figure}

In this part, 
we review the problems considered in \cite{comp1} and \cite{comp2}, where
we studied a compression design problem with a privacy constraint. Random variable $Y$ denotes the useful data and is correlated with the private data described by RV $X$. An encoder wants to compress $Y$ and communicates it to a user through an unsecured channel. The encoded message is described by RV $\mathcal{C}$. As shown in Fig.~\ref{ITWsys1}, it is assumed that an adversary has access to the encoded message $\mathcal{C}$, and wants to extract information about $X$. Moreover, it is assumed that the encoder and decoder have access to a shared secret key denoted by RV $W$ with size $M$. The goal is to design encoded message $\mathcal {C}$, which compresses $Y$, using a variable length code with minimum possible average length that satisfies certain privacy constraints. To solve the problems, we utilize techniques used in privacy mechanism in \cite{shah} and \cite{kostala} and compression design problems and combine them to build such $\mathcal {C}$. We divide the main results into two main parts: 1. Non-zero leakage; 2. Zero leakage.\\
\textbf{Part I (Non-zero Leakage):}
Here, the results in \cite{comp2} are summarized.
We have extended previous existing results \cite{kostala,kostala2}, by generalizing the perfect privacy constraint and allowing non-zero leakage. We have used different privacy leakage constraints, e.g., the mutual information between $X$ and $\mathcal{C}$ equals to $\epsilon$, strong privacy constraint and bounded per-letter privacy criterion. An approach is introduced and is applied to a lossless data compression problem. \\ 
\textbf{Part II (Zero Leakage):}
In this part, the results in \cite{comp1} are summarized.
Considering two different cases we have extended previous existing results \cite{kostala,kostala2} and improved the bounds. 
Here, we considered the problem in \cite{kostala} where a private compression design problem with zero leakage is studied and we have extended the results under the following assumptions: 1. The realization of $(X,Y)$ follows a specific joint distribution; 2. $|\mathcal{X}|\geq |\mathcal{Y}|$.
To this end we combined the privacy design techniques used in \cite{shah}, which are based on extended versions of Functional Representation Lemma (FRL) and Strong Functional Representation Lemma (SFRL), as well as the lossless data compression design in \cite{kostala}. We obtained lower and upper bounds on the average length of the encoded message $\mathcal{C}$ and evaluated them in different scenarios.
Considering a specific set of joint distributions for $(X,Y)$, we proposed an algorithm with low complexity to find bounds on the optimizer of privacy problems in \cite{shah} with zero leakage. We applied the obtained results to design $\mathcal{C}$ which achieves tighter bounds compared to \cite{kostala}. Furthermore, when $|\mathcal{X}|\geq |\mathcal{Y}|$, we proposed a new code that improves the bounds in \cite{kostala}. 
Finally, in a numerical example we evaluated the bounds and compare them with \cite{kostala}.
\subsection{Semantic communications with privacy constraints}
\begin{figure}[]
	\centering
	\includegraphics[scale = .53]{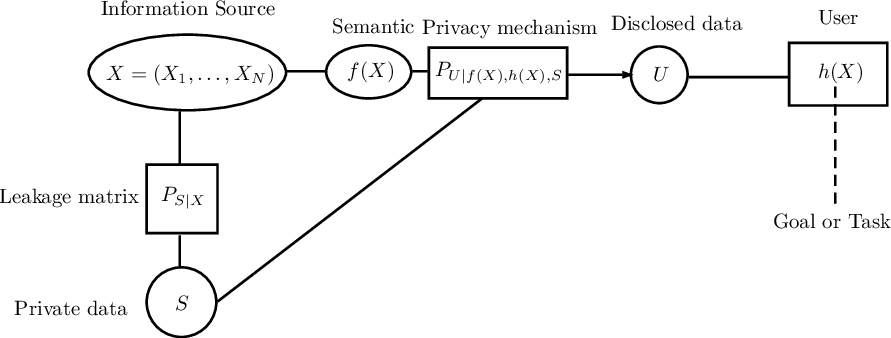}
	\caption{Private semantic communication model. The goal is to design $U$ such that it keeps as much information as possible about $h(X)$ while satisfying a privacy constraint.}
	\label{semsysch11}
\end{figure}
Here, we review the results in \cite{zamani2024semantic}.
In this paper, random variable (RV) $X=(X_1,\ldots,X_N)$ denotes the information source and is correlated with the private data denoted by RV $S$. 
As shown in Fig.~\ref{semsysch11} a user asks an encoder about a task denoted by a function of $X$, i.e., $h(X)$. In this work $h(X)$ describes the task or goal of the communication. The encoder designs a message which is a function of $X$, i.e., $f(X)$, to disclose it. Here, $f(X)$ corresponds to the semantic of the information source which has less dimension compared to $X$. Since $f(X)$ can not be revealed directly (due to the privacy constraints) the encoder utilizes a privacy mechanism to produce disclosed data described by RV $U$. The goal is to design $U$ based on the goal, semantic, and private data that reveals as much information as possible about $h(X)$ and satisfies a privacy criterion. We use mutual information to measure utility and privacy leakage. In this work, some bounded privacy leakage is allowed, i.e., $I(S;U)\leq \epsilon$. In this problem, we utilized concepts from the privacy mechanism design outlined in \cite{shah} and \cite{zamanistatistical}
to introduce an innovative private semantic communication framework. The proposed scheme offers a mathematical approach to design a goal-oriented private utility function. This function not only facilitates the receiver in achieving the goal but also guarantees the privacy of the data from the recipient.
For the privacy mechanism design which corresponds to the acheivabilty we used different methods. To this end, extended versions of the Functional Representation Lemma and the Strong Functional Representation Lemma and separation technique are used to address a \emph{private semantic communication} problem. To produce disclosed data $U$ an approach is proposed where artificial noise denoted by RV $M$ is added to the semantic. We then use the privacy mechanism to design the artificial noise. In this work we assumed that both semantic and goal are known by the encoder. Lower and upper bounds on the privacy-utility trade-off are provided and the bounds are studied in special cases.
\vspace{-4mm}
\subsection{Multi-task semantic communications with privacy constraints}
\begin{figure}[]
	\centering
	\includegraphics[scale = .48]{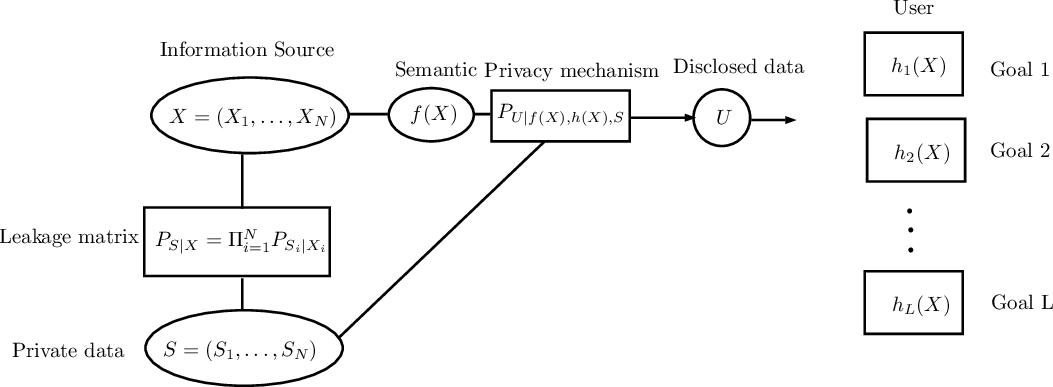}
	\caption{Multi-task private semantic communication model. The goal is to design disclosed data $U$ such that it maximizes weighted linear combination of utilities while satisfying a certain privacy constraint. In this model, we assume that each task is a subset of the information source $X$, i.e., $h_i(X)\subseteq \{X_1,\ldots,X_N\}, \ \forall i$.}
	\label{semsysch1122}
\end{figure}
Here, we summarize the results in \cite{Multitask}. As shown in Fig.~\ref{semsysch1122}, we consider a multi-task communication scenario, where the information source $X=(X_1,\ldots,X_N)$ is correlated with the private data $S=(S_1,\ldots,S_N)$ through the leakage matrix $P_{S|X}=\Pi_{i=1}^{N}P_{S_i|X_i}$, and $S_i$ and $X_i$ are arbitrary correlated. A use asks for $L$ tasks $h(X)=(h_1(X),\ldots,h_L(X))$, where each task is a subset of the information source, i.e., $h_i(X)\subseteq \{X_1,\ldots,X_N\}$. The encoder designs the semantic that maximizes the a linear combination of the utilities based on the tasks. The utility of each task is measured by the mutual information between the task and the semantic. Similarly, due to the privacy constraints the encoder utilizes a privacy mechanism to produce disclosed data described by RV $U$. The disclosed data $U$ is designed such that it maximizes weighted linear combination of utilities while satisfying a privacy criterion. We use mutual information to measure the privacy leakage, i.e., $I(S;U)\leq \epsilon$. Here, we use concepts from the privacy mechanism design outlined in \cite{shah} to introduce novel single and multi-task private semantic communication frameworks. In this paper, to design the disclosed data different methods are considered. Extended versions of the Functional Representation Lemma and the Strong Functional Representation Lemma, as well as a separation technique are used to address the \emph{multi-task private semantic communication} problem. Privacy preserving message is obtained by adding noise to the semantic information. A design is proposed that maximizes a weighted linear combination of the utilities achieved by the user while satisfying a privacy constraint. It has been shown that the highly challenging and complex problem can be decomposed into independent single-task problems and a simple privacy mechanism design can be obtained.
In contrast with previous works on goal oriented semantic communication, which are mostly based on deep learning, in this paper a simple theoretical framework is proposed to design the disclosed data considering multi-task private semantic communication problems.
\vspace{-5mm}
\subsection{Fair and private representations}
\begin{figure}[t]
	\centering
	\includegraphics[scale = .5]{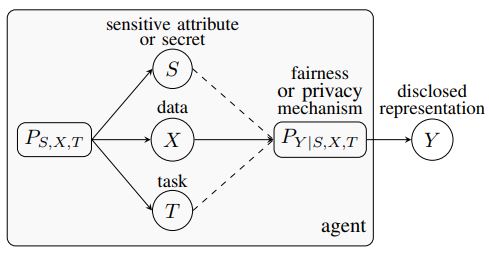}
	\caption{Data representation with perfect demographic parity or privacy. We want to design a representation $Y$ of the data $X$ that is useful for the task $T$, is compressed, and independent of the sensitive attribute or secret $S$.}
	\label{fig:ISITsys55}
\end{figure}
Here, we review the problem in \cite{zamani2024information}. In this article, we studied the fundamental limits in the design of fair and/or private representations achieving perfect demographic parity and/or perfect privacy through the lens of information theory. More precisely, given some useful data $X$ that we wish to employ to solve a task $T$, we consider the design of a representation $Y$ that has no information of some sensitive attribute or secret $S$, that is, such that $I(Y;S) = 0$. We consider two scenarios. First, we consider a design desiderata where we want to maximize the information $I(Y;T)$ that the representation contains about the task, while constraining the level of compression (or encoding rate), that is, ensuring that $I(Y;X) \leq r$. Second, inspired by the Conditional Fairness Bottleneck problem, we consider a design desiderata where we want to maximize the information $I(Y;T|S)$ that the representation contains about the task which is not shared by the sensitive attribute or secret, while constraining the amount of irrelevant information, that is, ensuring that $I(Y;X|T,S) \leq r$.
\section{future directions}
As we outlined earlier, the approach proposed by \cite{kostala} and \cite{shah}, where in \cite{shah} we have extended the FRL and the SFRL, can be applied to a variety of information theory and communication problems. 
For future works, one could extend the FRL and SFRL by considering different measures. For instance, these extensions could be used to address privacy problems where differential privacy is the leakage measure. One could also consider other privacy leakage measures such as maximal leakage. Moreover, we have combined information geometry concepts with extended versions of the FRL and the SFRL to address the privacy problems with point-wise measures. One can use different point-wise measures like the lift \cite{e25040679}.  
\vspace{-5mm}
\section{conclusion}
In this paper, we reviewed the approach proposed by \cite{shah}, where we extended the Functional Representation Lemma (EFRL) and the Strong Functional Representation Lemma (ESFRL) to develop privacy mechanism designs. It is argued that, due to their constructive proofs, these lemmas are valuable for privacy design. The proposed mechanism designs are optimal under specific constraints, which can be generalized using the concept of common information. Additionally, this approach has been shown to apply to various information-theoretic problems, such as semantic communications, compression design, and caching and delivery with privacy constraints. Furthermore, we discussed how concepts from information geometry can be combined with this approach to address privacy mechanism design problems with point-wise privacy leakage measures. A notable aspect of point-wise privacy measures is the potential for geometric interpretations of the resulting designs. Moreover, point-wise measures are stronger than average-based privacy leakage measures, as they must hold for each alphabet of the disclosed data.
\newpage
\bibliographystyle{IEEEtran}
{\balance \bibliography{IEEEabrv,IZS}}
\end{document}